\documentclass[twocolumn]{aa}
\usepackage{graphicx}
\usepackage{txfonts}
\def\Msol{\thinspace\hbox{$\hbox{M}_{\odot}$}}

\def\a4{\hsize 17.0cm \vsize 25.cm}

\begin{document}
\title{On the neutral gas distribution and kinematics in 
the dwarf irregular galaxy IC 1613}

\titlerunning{The neutral gas kinematics in IC 1613}

\author{Sergiy Silich \inst{1}
        \and
        Tatyana Lozinskaya\inst{2}
          \and 
        Alexei Moiseev \inst{3}
          \and 
        Nikolai Podorvanuk  \inst{2}
          \and
        Margarita Rosado  \inst{4}
          \and
       Jura Borissova  \inst{5}
          \and
        Margarita Valdez-Gutierrez \inst{6}
}

\offprints{S. Silich}

\institute{Instituto Nacional de Astrof\'\i sica Optica y
                 Electr\'onica, AP 51, 72000 Puebla, M\'exico\\ 
                 \email{silich@inaoep.mx}
         \and
Sternberg Astronomical Institute, Universitetskii pr. 13,
                 Moscow, 119992 Russia\\ 
                 \email{lozinsk@sai.msu.ru}
         \and
Special Astrophysical Observatory, Russian Academy of
               Sciences, Nizhnii Arkhyz, 357147 Karachai-Cherkessian 
               Republic, Russia\\
               \email{moisav@sao.ru}
         \and
Instituto de Astronom\'\i a, Universidad Nacional
               Aut\'onoma de M\'exico, AP 70-264, 04510 M\'exico DF, 
               M\'exico\\
               \email{margarit@astroscu.unam.mx}
         \and
European Southern Observatory, Ave., Alonso de Cordova 3107,
               Casilla 19 Santiago 19001, Chile\\
               \email{jborisso@eso.org}
         \and
Instituto de Astronom\'\i a, UNAM,  
             AP 877, Ensenada 22800, B.C., M\'exico\\
               \email{mago@astrosen.unam.mx}
             }

   \date{Received April 28, 2005; accepted October 13, 2005}

\abstract{
{\it Aims.} We study the neutral hydrogen distribution and
kinematics in the Local Group dwarf irregular galaxy IC 1613
and compare them with the ionized gas distribution and stellar
content of the galaxy. We discuss several mechanisms which
may be responsible for the origin of the observed complicated
HI morphology and compare parameters of the most prominent 
kpc-scale HI structure with the multiple SNe scenario.

\noindent
{\it Methods.} The observations were performed with the Vary Large
Array of NRAO with a linear resolution $\sim 23$ pc at the
adopted distance of 725 kpc and the spectral channel width of 2.57
km s$^{-1}$. The numerical calculations have been provided with
our 2.5D Lagrangian scheme based on the thin layer approximation.

\noindent
{\it Results.} We found that the ISM of the galaxy is highly
inhomogeneous and identified  a number of intermediate-scale 
(200 pc - 300 pc in diameter) HI arcs and shells having expansion 
velocities of 10 to 20 km s$^{-1}$. Besides these shells, several 
giant holes and arc-shaped structures have been revealed, 
whose radii exceed several hundred parsecs. We found that 
parameters of the most prominent ($M_{HI} = 2.8 \times 10^7$\Msol)
kpc-scale structure and the level of the detected star formation 
activity are inconsistent with the multiple SNe hypothesis.
   }

   \maketitle
%

\section{Introduction}

The neutral component of the ISM in star-forming dwarf and spiral 
galaxies presents a collection of holes, arcs, and shell-like structures 
(Davies \& Tovmassian, 1963; Heiles, 1984;  Brinks \& Bajaja, 1986; 
Puche et al. 1992; Ehlerov\'a et al. 2004). Recent studies of late-type dwarf 
galaxies indicate that $\sim 1/3$ of the galaxies present a high column 
density and sometimes broken rings with $N_{HI} > 10^{21}$ cm$^{-2}$
(Stil \& Israel, 2002).  The origin of HI holes and shell-like 
structures has been the subject of debate for more than two
decades, but still remains a controversial issue (see, for example, Block
\& Walter, 2005).

Several different mechanisms have been proposed for
the formation of the observed HI distribution and kinematics in
different galaxies (see, for example, reviews by Tenorio-Tagle \&
Bodenheimer, 1988; van der Hulst, 1996).

In the standard approach based on the Dyson \& de Vries (1972), 
Weaver et al. (1977) model, HI shells result from the cumulative 
effects of multiple stellar winds and supernovae explosions inside 
the volume presently encompassed by the shell. In such 
a case the energy released by stellar winds and
supernovae is thermalized and generates a hot, high pressure cavity
that drives a shock wave into the surrounding interstellar medium (ISM).
This shock collects interstellar gas and compresses it into a dense,
expanding shell. As the hot cavity continues to expand, the
temperature and pressure inside the cavity drop while the outer shell
decelerates. When the
pressure becomes comparable to that of the ambient medium, the shock
wave vanishes. At such a pressure-confined stage (Koo \& McKee, 1992),
the shell stalls and starts to disintegrate with its local sound
speed. 

Despite many cases in which this approach gives a plausible explanation
for the observed structures (see, for example, 
Oey \& Garc\'{\i}a-Segura, 2004; Vorobyov et al., 2004,
2005; van der Hulst, 1996 and references therein), it was recognized long ago
(Heiles, 1984; Tenorio-Tagle \& Bodenheimer, 1988) that this scenario
cannot explain all types of observed structures. In particular, it
encounters problems when attempting to explain the origin of the largest
supershells where the stellar associations found within the HI rings
seem to be inadequate energy sources. More recently, Rhode et al. (1999)
found that in the case of the Holmberg II (HoII) galaxy, the observed 
upper limits for 
stellar cluster remnants detected inside neutral hydrogen holes is in
many cases inconsistent with the SNe hypothesis. Although a
year later Stewart et al. (2000) re-examined their results using deep
far-ultraviolet (FUV), H$\alpha$ and HI data and decided that the energy 
deposited by stellar winds and SNe explosions into the ISM of HoII is 
sufficient, there are many HI holes that have no optical counterparts
consistent with the rate of mechanical energy required by the
standard model.

In other galaxies, Kim et al. (1999) found only a weak correlation
between the LMC neutral hydrogen holes and the HII regions and
concluded that the hypothesis of multiple winds and supernovae is 
inconsistent with their data. 
Perna \& Gaensler (2004) proposed a non-traditional approach to the
problem by counting the number of radio pulsars associated with Milky Way
supershells. They performed Monte-Carlo simulations for the expected
pulsar population and predicted that several tens of radio pulsars may
be detected in the directions to the largest Milky Way supershells, and
few of them should be found inside regions encompassed by the contours of 
smaller shells. They found that for the smaller shells their predictions 
are consistent with the detected pulsar population. However for the giant 
supershell GSH242-03+37 the observed number of radio pulsars is inconsistent 
with the multiple winds and supernovae model. 

Several other mechanisms that do not require violent stellar activity
include collisions of high velocity clouds with galactic disks 
(Tenorio-Tagle, 1981), the radiation pressure from field stars
(Elmegreen \& Chiang, 1982), the non-linear evolution of self-gravitating
turbulent galactic disks (Wada et al. 2000; Dib \& Burkert, 2004), the 
ram pressure of the intergalactic medium (Bureau \& Carignan, 2002), 
and even exotic mechanisms such as the distortion of the ISM by  
powerful gamma-ray bursts (Efremov et al. 1999).
Despite many efforts attempting to resolve this problem, the detailed 
comparison of different scenarios with objects detected in different
galaxies remain scarce and controversial.

IC 1613 is a member of the Local Group and is located 725 - 730 kpc 
from the Milky Way (Freedman, 1988a,b; Dolphin et al., 2001). Because 
of its proximity , it represents an excellent target to study the 
interplay among the stellar and gaseous components in galaxies.
Its HI mass, derived from  single-dish 
observations by Lake \& Skillman (1989), is $6 \times 10^7$\Msol. 
The galactic rotation is slow, with an almost linear rotation curve
inside the inner 2.5 kpc radius, having a maximum amplitude of rotation 
velocity around 25 km s$^{-1}$ (Lake \& Skillman, 1989). 

The stellar population of IC 1613 and its star formation history (SFH)
have been studied by a number of authors. Hunter et al. (1993)
estimated the total H$\alpha$ luminosity to be $L_{H\alpha} \approx 3 \times
10^{38}$ erg s$^{-1}$. The associated star formation rate (SFR) is in
good agreement with that obtained by Cole et al. (1999), who found 
SFR$ \approx 3 \times 10^{-3}$ \Msol \, yr$^{-1}$ over the past 10
Myr. Skillman et al. (2003) did not find any evidence for recent
episodes of violent star formation in this galaxy from their analysis
of color-magnitude diagrams and concluded that it looks unlikely that
this and similar galaxies may be responsible for the chemical enrichment
of the intergalactic medium. Georgiev et al. (1999) obtained deep
UBV photometry for more than 3000 stars in selected areas of IC
1613. The isochrone fit of their data clearly indicates the
presence of young (age $< 30$ Myr) stars and stellar associations.

On the other hand, Fabry-Perot interferometric studies of the
ionized gas kinematics (Valdez-Guti\'errez et al. 2001; Lozinskaya
et al. 2003) revealed a number of supershells covering the whole extent 
of the optical galaxy. The complex radial velocity pattern indicates
expansion velocities up to 60 -- 75 km s$^{-1}$. The majority of 
the supershells presents an interior OB-association, suggesting a
physical link between them. 
Lozinskaya et al. (2003) identified three large ($\sim 300$ pc), bright 
HI supershells in the north-east part of the galaxy. They
expand at $\sim 15$ km s$^{-1}$ and overlap with several smaller 
HII shells distributed along the walls of the HI supershells and are 
ionized by the young OB-associations. The location of the HII 
shells and OB-associations suggests that sequential star formation 
was triggered in this region by larger neutral hydrogen supershells.
This implies that, despite a quiet history of star formation, the
ISM of IC 1613 has complex gas kinematics and morphology which 
are linked to the embedded stellar population.

Here we present a study of the neutral hydrogen
distribution and kinematics within this galaxy based on Very 
Large Array observations and compare them with our previous studies of
the ionized gas and stellar content in this galaxy. We have found 
several giant HI rings and arc-like structures. The smallest ones 
present clear 21-cm line splitting indicating that in these cases 
we observe expanding shells generated by the embedded
OB-associations. The origin of the largest ones is not clear.
In particular, the largest, $\sim 1$ kpc in diameter, structure
contains several OB-associations of different ages which appear to be
potential sources of energy. However we did not find 
any indication of a regular shell expansion that leads to an
assumption that this structure may represent a case of the standing,
pressure-confined shell. This initiated us to compare the observed
parameters of the ``main supershell'' with the standard
model based on the multiple winds and supernovae hypothesis.
We found that the SNe hypothesis is in poor agreement
with the observationally restricted  rate of star formation that 
indicates a different origin of this large, kpc-scale structure. 

The paper is organized as follows. In Section 2 we present the results
of the observations, the procedures for the data reduction, discuss
the neutral hydrogen distribution and the observed velocity field and
compare this with properties of the ionized ISM and stellar content
of the galaxy. In particular, we discuss properties of the largest 
HI structure 1 kpc in diameter that is similar to the largest HI
supershells detected in other galaxies. In Section 3 we discuss the
standard superbubble model, present
the results of numerical calculations and compare them with the 
observed parameters of the ``main supershell'', the largest 
HI structure detected in IC 1613. Section 4 summarizes our results.

\section{HI data: morphology, kinematics and their links 
with other galactic components}

Neutral hydrogen observations were made with the Very Large Array
(VLA) of The National Radio Astronomy Observatory (NRAO) in 
configurations B, C and D. The application for the current  
observations was compiled by E.M. Wilcots. The first results, together
with the brief discussion for data reduction, were presented in 
Lozinskaya et al. (2001).

The data cube has an angular resolution of $7^{\prime\prime}.4 \times
7^{\prime\prime}.0$, and a pixel size of $3^{\prime\prime} \times 
3^{\prime\prime}$. This implies a linear resolution
of~${\sim23}$~pc at the adopted distance of 725~kpc. The spectral 
channel width is 2.57~km~s$^{-1}$. Our data cube allows us 
to derive the distribution and kinematics of the HI gas of the 
entire galaxy. To build up the position-velocity diagrams and the
integrated 21-cm line intensity map, the 40 velocity channels, from 
the 127 available, have been selected by the condition that the 
signal-to-noise ratio exceeds the $2\sigma$ limit in three or 
more adjacent velocity channels (a windowing technique; see, for example,
Thornley \&  Mundy, 1997). The selected 
channels cover velocity interval from $-279$ to $-178$ km s$^{-1}$.

\subsection{The distribution of neutral hydrogen}

The integrated 21-cm line flux density derived from our data is
$482\,\mbox{Jy}\,\mbox{km}\,\mbox{s}^{-1}$. This implies that the total 
HI mass of the galaxy is $6.0 \times 10^7\,M_\odot$. This value is in good 
agreement with the single-dish measurements by Lake \& Skillman (1989;
see Table~II from their paper), despite a ``negative bowl" in our data 
that leads to a negligible underestimation of the HI mass. An average 
1D velocity dispersion of  5.9~km~s$^{-1}$ was measured from our data 
cube after accounting for the instrumental resolution. This corresponds
to a 3D velocity dispersion $\sigma_{ISM} = 10.2$~km~s$^{-1}$. 
Lake \& Skillman (1989) adopted a somewhat larger value of 7.5 km~s$^{-1}$ 
for the 1D velocity dispersion. However, our data have better spectral 
and spatial resolutions (5 km~s$^{-1}$ versus 6.2 km~s$^{-1}$, and 
7$^{\prime\prime}$ versus 45$^{\prime\prime}$).

The map of the 21-cm line surface brightness is presented in Figure 1a. 
The three brightest and most clearly defined shells (Shells I, II and
III) in the N-E sector of the galaxy have diameters of 200 - 300 pc and are 
clearly related to the only region of noticeably enhanced star formation
discovered so far in the galaxy (see Lozinskaya et al. 2003).
Several more shells and arcs of similar sizes having lower surface 
brightness are clearly identified in Figure 1a and Figure 3a
(e.g. Shells IV and V).
\begin{figure*}
\vspace{20.0cm}
\centering
\includegraphics{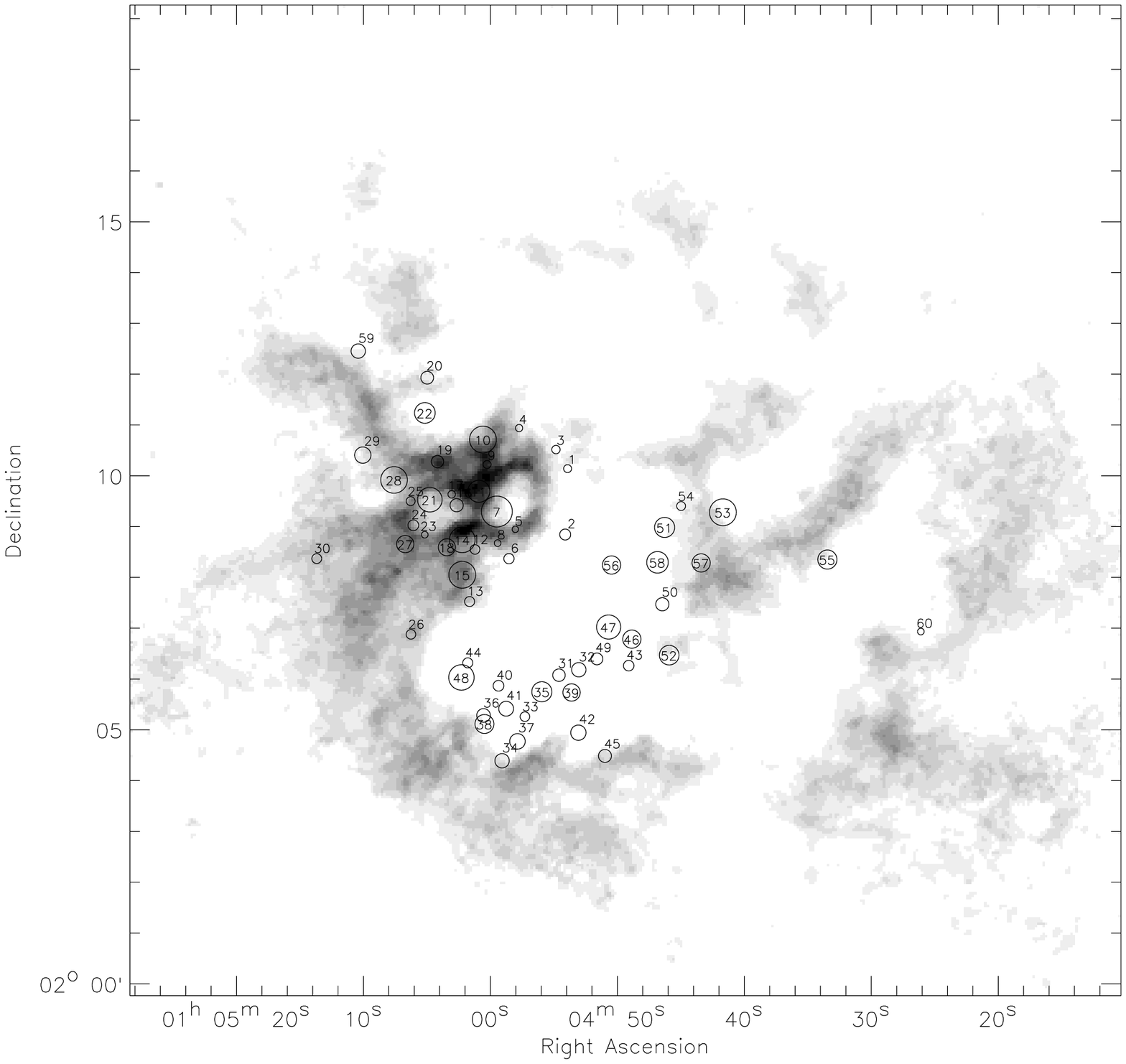}
\includegraphics{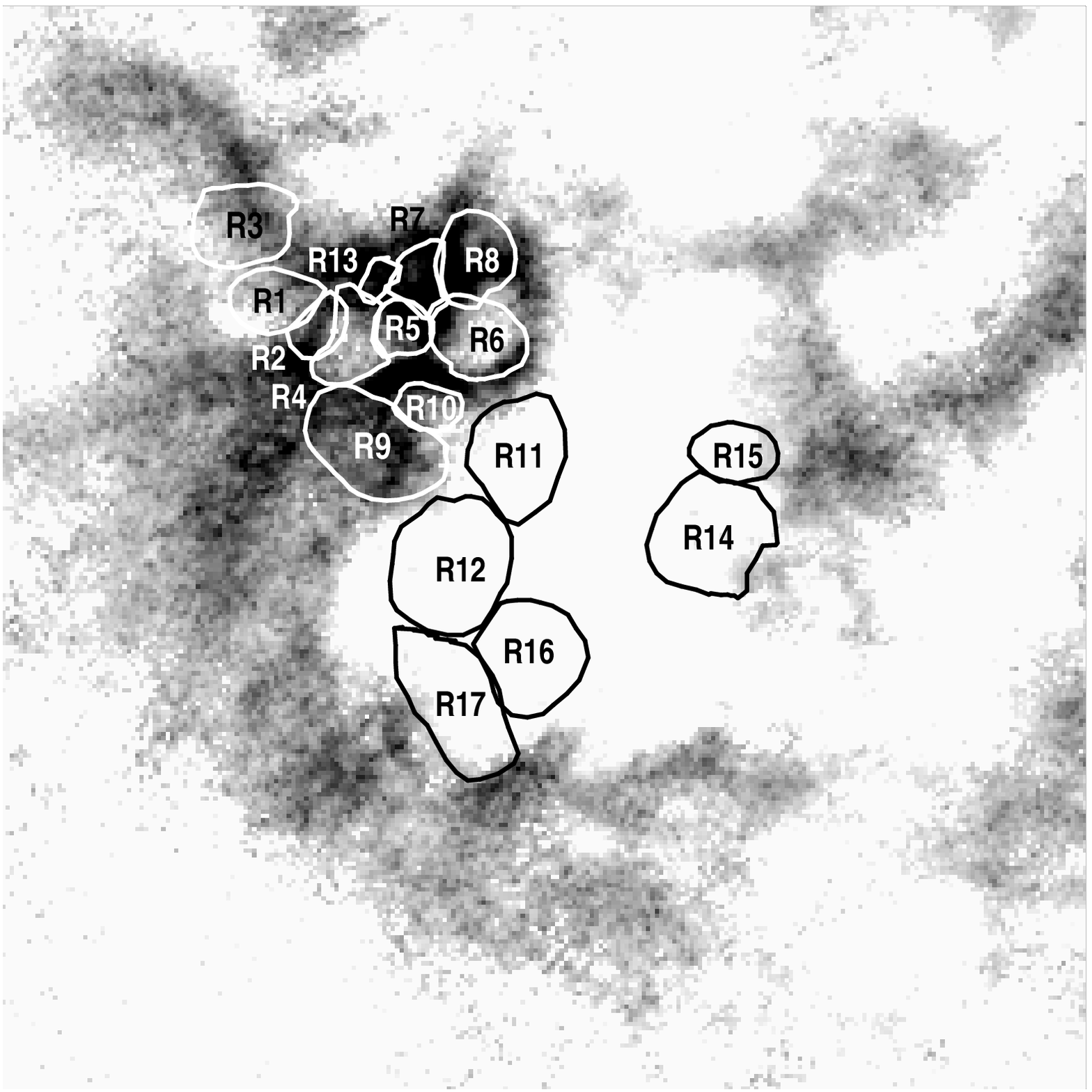}
\includegraphics{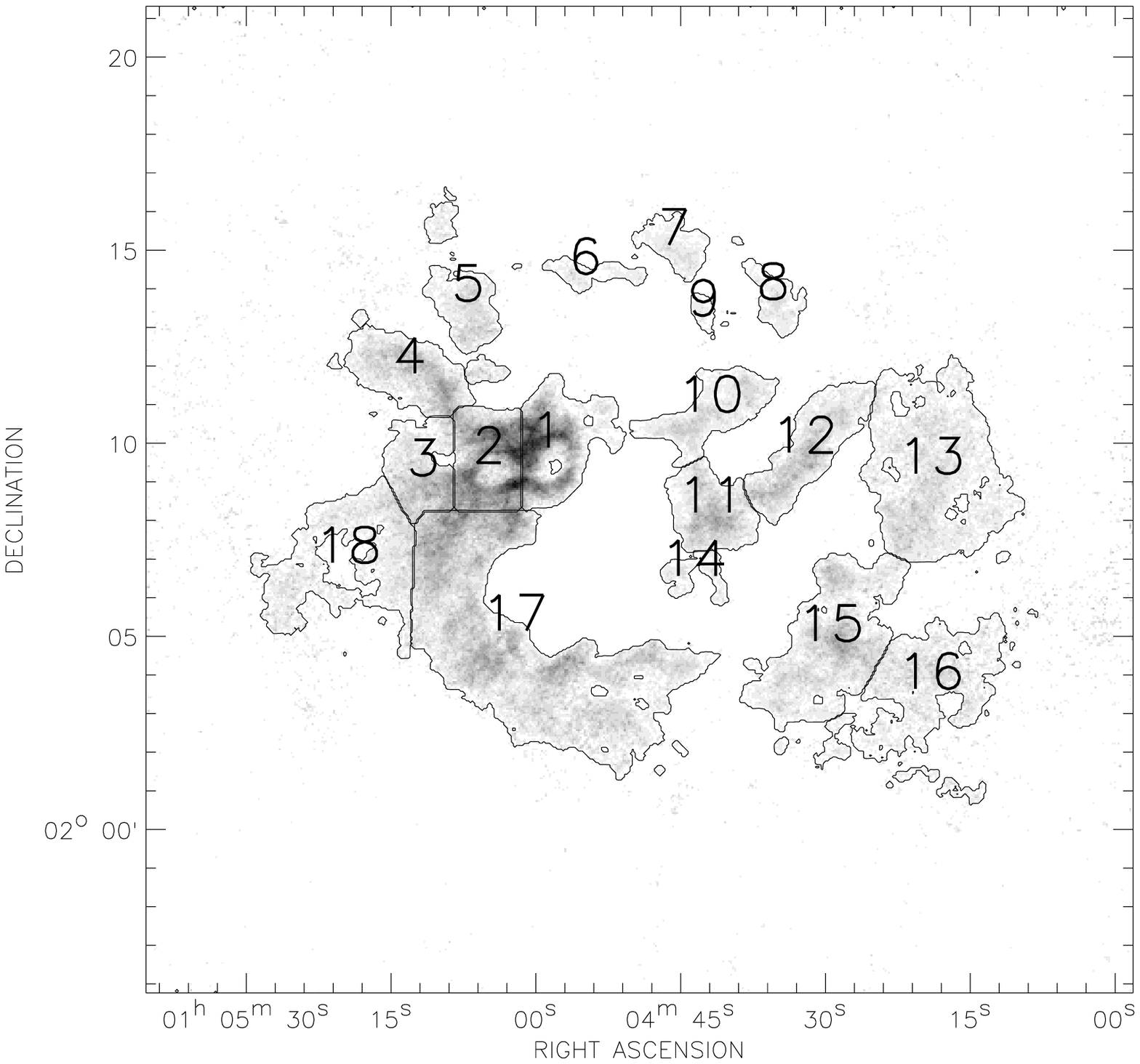}
\caption{The distribution of neutral hydrogen. Upper left: the 21-cm 
line brightness map with OB associations from the 
list of Borissova et al. (2004) superposed. OB associations are shown 
by circles whose radii are proportional to their sizes. The
coordinates and radii of associations are taken from the Table 1 of 
Borissova et al. (2004). Upper right: the 21-cm line brightness
map of the southeast sector of the galaxy with the 
superbubbles of Valdez-Guti\'errez et al. (2001) superposed. Lower
panel: the distribution of the HI column density throughout the galaxy 
split into a number of segments. In all panels north is at the top 
and east is on the left.} 
\label{fig1}
\end{figure*}

Besides these intermediate-size shells, one can easily recognize in
Figure 1 several larger arcs and ring-like structures which are 
similar in size and morphology to the supergiant shells revealed by 
Meaburn (1980) at the Large Magellanic Cloud.
The most prominent structure seen in Figure 1a coincides with the giant 
region of low surface brightness emission visible in the lower angular 
resolution map of Lake and Skillman (1989). It is a hole 
surrounded by a thick HI ridge. The hole has a 5$^{\prime}$ diameter 
($\sim 1$ kpc diameter at the distance adopted for IC 1613) and is 
centered at RA(2000) = 1h 4m 52s, Dec(2000) = 2$^\circ 07^{\prime}$. 
The thickness of the HI ridge, $\Delta R$, is about 200 pc. The
Northern, Eastern and Southern parts of the ring   
have a complicated morphology and a kinematical pattern suggesting a feedback 
from the local star formation. Shells II and III form the brightest part of 
the northern wall. The northwest sector of the ring appears to have a
lower surface brightness and is not clearly identified in our map. 
Figure 1a also presents the locations of the OB-associations (shown by 
circles whose radii are proportional to the sizes of the
OB-associations) taken from the list of Borissova et al. (2004). 
Figure 1b displays a population of superbubbles identified 
by Valdez-Guti\'errez et al. (2001). Hereafter the whole structure 
is referred to as the ``main supershell''.

We split the galaxy into a number of segments (Figure 1c) and then
calculated the mass of each segment by integration of the HI column 
density over the segment surface. The segment masses are presented in 
Table 1. We associate the ``main supershell'' with segments 1, 2, 11, 14, 
17 (Figure 1c). The HI mass of the ``main supershell'' is then $2.8 \times 
10^7$ \Msol, which corresponds to a total mass of $3.9 \times 
10^7$ \Msol \, when including helium.
\begin{table}
   \label{tab1}
\caption{The distribution of the HI mass in the galaxy.}
\begin{center}
{\small
\begin{tabular}{|l|c|c|c|c|c|c|}
\hline
Segment & 1 & 2 & 3 & 4 & 5 & 6 \\   
\hline\hline  
Mass ($10^{6}M\odot$) & 5.37 & 6.17 & 2.39 & 2.86 & 1.70 & 0.38   \\
\hline
Segment & 7 & 8 & 9 & 10 & 11 & 12 \\  
\hline 
Mass ($10^{6}M\odot$) & 0.57 & 0.48 & 0.14 & 1.57 & 2.20 & 2.66 \\
\hline
Segment & 13 & 14 & 15 & 16 & 17 & 18 \\  
\hline 
Mass ($10^{6}M\odot$) & 4.95 & 0.34 & 4.46 & 3.60 & 14.02 & 3.18 \\ 
\hline
\multicolumn{7}{|l|}{Total mass ($10^{6}M\odot$): 57.0} \\
\hline
\end{tabular}}
\end{center}
\end{table}

Another example of a large shell-like structure is an elliptical
hole in the HI distribution at the north of the main supershell 
surrounded by sectors 1, 5, 6, 7, 9 and 10 (Figure 1c). One can also 
recognize several incomplete shells formed by sectors 10, 11 and 12, 
sectors 14, 11, 12, 13 and 15, and sectors 13, 15 and 16 at the northwest
and west of the main supershell. The size of the shell to the north 
is similar to that of the ``main supershell'' but it has a smaller
mass, $M_{HI} \approx 9.7 \times 10^6$ \Msol.

\subsection{The ionized component, the stellar population, and the local SFR}

The examination of the distribution of the HII regions and superbubbles 
from the list of Valdez-Guti\'errez et al. (2001; Tables 4 and 5) shows 
that many of them are located inside the area encompassed by the contour 
of the ``main supershell'' or directly inside the HI ridge.
Figure 1b displays the boundaries of the superbubbles found by 
Valdez-Guti\'errez et al. (2001) superimposed on the HI image
of the galaxy. Several superbubbles (R11, R12, R14, R15, R16 and R17)
are located inside the ``main supershell''. R1, R4 and R6 seem to coincide
with HI shells I, II and III identified by Lozinskaya et al. (2003).
The rest are projected onto the HI ridge and seem to be associated
with the extended region of star formation at the northeast of the galaxy.

The $H{\alpha}$ luminosities of the HII regions and superbubbles
obtained by Valdez-Guti\'errez et al. (2001) allow us to estimate the
mass of the ionized gas and the local star formation rate 
within the area encircled by the ``main supershell''. The mass of the 
ionized gas associated with superbubbles was calculated from the 
relation $M_{HII} = 3.3 \times 10^{-2} \sum (R_i / 1 pc)^3 
\, n_{e,i}$ (\Msol), where radii of superbubbles, $R_i$, and electron 
number densities, $n_{e,i}$, were taken from Table 8 of 
Valdez-Guti\'errez et al. (2001), the thickness of the superbubble 
shells was assumed to be 1/12 of their radii, and the contribution
from the helium was taken into account. For the HII regions we assume a 
homogeneous density distribution. Then $M_{HII} = 3.3 \times 10^{-1} \sum
(R_i / 1 pc)^3 \, n_{e,i}$. The rms electron number densities were
calculated from the relation $n_{e,i} = (2.3 \times 10^{17} F_{H\alpha,i} / 
\Theta_i^2 R_i)^{1/2}$, where $\Theta_i$ is the equivalent angular diameter 
of the HII region expressed in square arc-seconds and $F_{H\alpha,i}$ is the
$H{\alpha}$ flux in the units of ergs cm$^{-2}$ s$^{-1}$. The radii of the
superbubbles and the HII regions were measured in parsecs.
The total mass of the ionized component,
$M_{HII} =  1.6 \times 10^6$ \Msol, is small in comparison with the
HI mass of the main supershell and, therefore, we neglected its
contribution in our calculations.
\begin{figure}
\resizebox{\hsize}{!}{\includegraphics{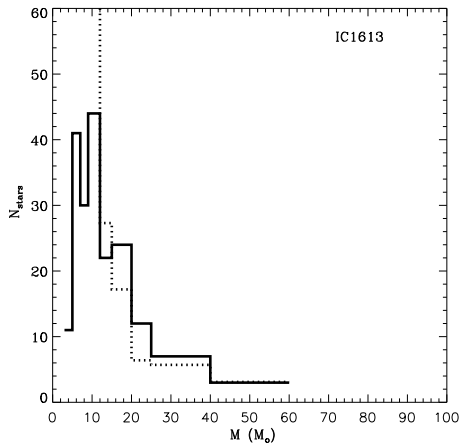}}
\caption{The local star formation rate. The number of massive stars, 
members of OB-associations, which are located inside the main supershell 
(solid line) as compared to the number of massive stars derived from 
the constant star formation rate model (dotted line). The reasonable
fit to the high mass top of the observed mass function requires the 
local rate of star formation to be $SFR_{local} = 4.5 \times 10^{-4}$ 
\Msol \, yr$^{-1}$.}
\label{fig2}
\end{figure}

We then estimated the local SFR using the calibration given in 
Kennicutt (1998) 
\begin{equation}
      \label{eq.0}
SFR(\Msol \, yr^{-1}) = 7.9 \times 10^{-42} L_{H\alpha} (erg \, s^{-1}) \,,
\end{equation}
adding up the luminosities of the superbubbles (R11, R12, R14, R15, R16 
and R17 from the list of Valdez-Guti\'errez et al. (2001), see Figure 1b)
that are located inside the inner boundary of the ``main supershell'' and 
therefore may contribute to the dynamics of the supershell. 
We found that the reddening correction is small because, according to 
the NASA/IPAC-Extragalactic-Database (NED), $A_v$ is only 0.083 mag.
The $H{\alpha}$ luminosity then is $3.24 \times 10^{37}$ erg s$^{-1}$, 
and the local star formation rate, $SFR_{local} = 2.6 \times 10^{-4}$ 
\Msol \, yr$^{-1}$. 
This value has been compared with the SFR (Figure 2) obtained from the 
analysis of stellar contents of OB-associations from the list of 
Borissova et al. (2004). The masses, the absolute visual magnitudes, M$_v$, 
and the ages of OB-associations within the field of the main supershell 
are listed in Table 2. 
\begin{table}
   \label{tab2}
     \caption{OB-associations from Borissova et al. (2004) that are
              located within the field of the main supershell}
\begin{center}
{\small
\begin{tabular}{|l||c|c||c||r|}
\hline
OB-association & Number of O,B stars & Mass    & M$_v$ & Age   \\
               &                     & (\Msol) &     & (Myr) \\
\hline\hline
31  & 6  &  62   &  -4.48  &  25 \\
32  & 7  &  107  &  -5.28  &  19 \\
33  & 10 &  109  &  -4.18  &  24 \\
34  & 4  &   35  &  -3.04  &  30 \\
35  & 17 &  236  &  -5.90  &  16 \\
36  & 7  &   47  &  -5.08  &  20 \\
37  & 7  &   59  &  -5.47  &  18 \\
38  & 21 &  176  &  -5.47  &  18 \\
39  & 6  &   48  &  -4.91  &  21 \\
40  & 4  &   25  &  -5.29  &  19 \\
41  & 5  &   35  &  -4.03  &  25 \\
42  & 6  &   46  &  -3.39  &  28 \\
43  & 4  &   17  &  -3.42  &  28 \\
44  & 5  &   40  &  -3.47  &  28 \\
45  & 5  &   34  &  -4.72  &  22 \\
46  & 7  &   60  &  -5.40  &  18 \\
47  & 7  &   41  &  -4.82  &  21 \\
49  & 4  &   19  &  -3.65  &  27 \\
50  & 4  &   22  &  -2.86  &  34 \\
51  & 5  &   35  &  -3.61  &  28 \\
52  & 6  &   36  &  -3.88  &  28 \\
53  & 7  &   52  &  -3.70  &  26 \\
54  & 6  &   28  &  -4.54  &  23 \\
55  & 6  &   25  &  -4.80  &  21 \\
56  & 4  &   28  &  -5.41  &  18 \\
57  & 4  &   14  &  -3.72  &  27 \\
58  & 7  &   38  &  -5.00  &  20 \\
\hline
\end{tabular}}
\end{center}
\end{table}

The ages of the associations have been calculated from the observed 
color-magnitude diagrams which were converted into  absolute
magnitudes and the true colors for each of the associations. 
The foreground reddening was calculated for each star individually 
using the reddening free $Q$ parameter and the latest calibration 
equations of Massey et al. (2000) summarized in their Table~4. A
distance modulus and a metallicity were adopted where 
$(m-M)_{0} = 24.27\pm0.1$ (Dolphin et al. 2000), and $Z=0.004$ 
(Mateo 1998). The ages were  then calculated by the best fit to the 
theoretical isochrones from the Geneva library (see Lejeune \& Schaerer
2001). The likely error of our method is 2-2.5 Myr.
 
In order to obtain the masses of individual stars and a generalized mass
function we have used the procedure proposed by Massey et al. (2000), 
which allows the determination of the effective temperatures ($T_{\rm eff}$) 
and the bolometric corrections (BC) for individual stars. The evolutionary 
tracks from Lejeune \& Schaerer (2001) for $Z = 0.004$ were then used to 
establish mass intervals. The local SFR derived from the comparison of 
this mass function with the constant star formation rate model is
$SFR_{local} = 4.5 \times 10^{-4}$ \Msol \, yr$^{-1}$. 

Obviously, these values only give a lower limit to the star formation
rate within the galactic area covered at the current epoch by the ``main 
supershell''. The mass function obtained from the Borissova et
al. (2004) data is incomplete at the low mass end and does not account 
for the field stars. The local SFR derived from the $H{\alpha}$ 
data is also related to the current epoch and does not consider the diffuse 
$H{\alpha}$ emission. Nevertheless our estimates are reasonable when 
compared to the mean star formation rate over the galaxy. Therefore we 
adopt them as a likely limit to the SFR within the field of the main 
supershell.

\subsection{The HI kinematics}

In order to study the kinematics of the interstellar medium (ISM) in 
IC 1613 we obtained position-velocity diagrams in more than 30 directions. 
The orientations of several scans and the corresponding position-velocity 
diagrams are presented in Figure 3.
\begin{figure*}
\vspace{22.0cm}
\centering
\includegraphics{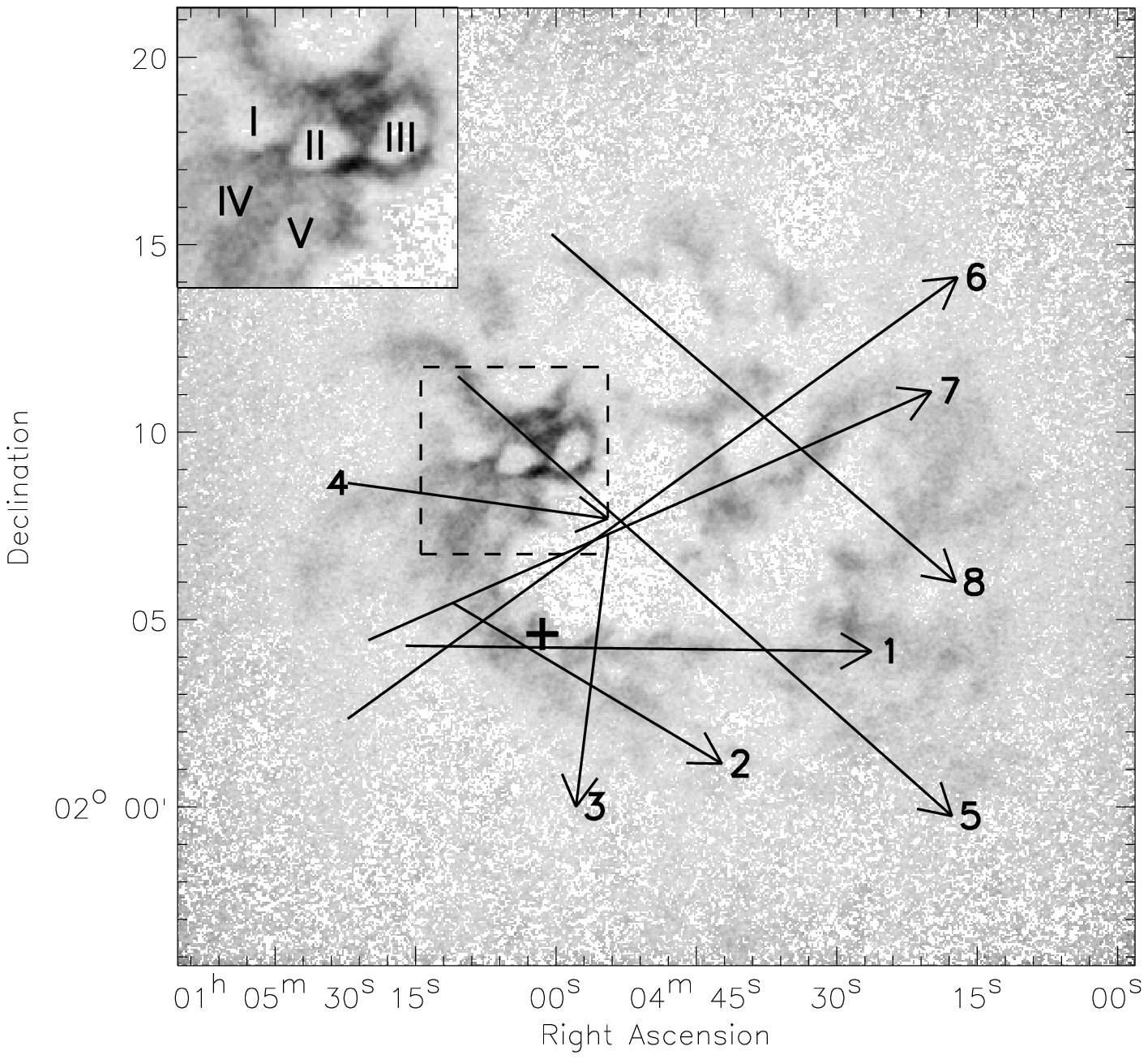}
\includegraphics{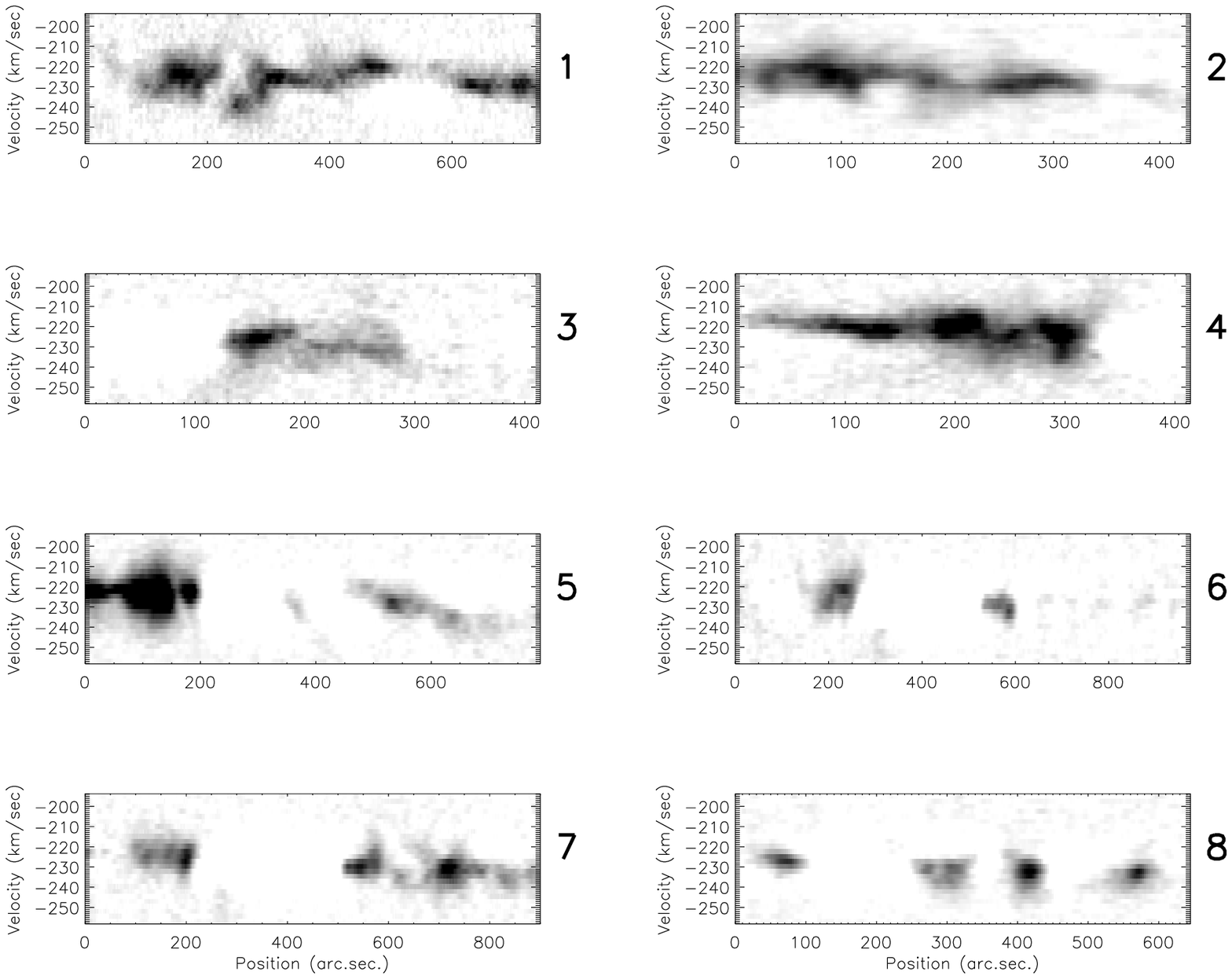}
\caption{The neutral gas kinematics. Upper panel displays orientations
of cuts used to study the neutral gas kinematics. Cuts are
identified by numbers. The region of active star formation in
the northeast sector of the galaxy is shown in the upper left-hand box.
The position of the WO star is marked by a cross.
Panels at the bottom represent the position-velocity diagrams along each
cut indicated at the upper panel. All velocities are heliocentric.}
\label{fig3}
\end{figure*}

The kinematics of Shells I, II and III were discussed by Lozinskaya 
et al. (2003). Shells II and III have expansion velocities of 12 - 18 
km~s$^{-1}$ and kinematical ages 5.3 - 5.6 Myrs. The expansion velocity 
of Shell I could not be measured,  while its size is similar to that of 
Shells II and III.
Shells IV and V mentioned in the previous section are much fainter and 
their properties are harder to determine than those of Shells I, II and III. 
Their radii are about 150 - 200 pc. The position-velocity diagram taken 
across the eastern part of the ``main supershell'' (cut 4 at Figure 3a)
reveals an offset of approximately of 15 km s$^{-1}$ between the 
regular rotation velocity of the galactic gas 
(which is in between -225 km s$^{-1}$ and -235 km s$^{-1}$ in this region) 
and the maximum values found in the field of the shells. This is 
-215 km s$^{-1}$ for shell IV (at 170$^{\prime \prime}$ - 
230$^{\prime \prime}$ along the cut), and 
-220 km s$^{-1}$ at the field of Shell V (at 250$^{\prime \prime}$ - 
315$^{\prime \prime}$ along the cut). 
These offsets together with the characteristic shapes of local velocity 
ellipses suggest that the expansion velocities of both of the shells are 
around 15 km s$^{-1}$.    

The position-velocity diagram taken along a shell formed by sectors 
10, 11 and 12 also indicates a regular shell expansion 
(see cuts 7 and 8 at Figure 3, at 550$^{\prime \prime}$ -
700$^{\prime \prime}$ and 300$^{\prime \prime}$ - 400$^{\prime
\prime}$ along the cuts). The expansion velocity is around 13 km s$^{-1}$. 
The size of this shell is 3$^{\prime}$ that corresponds to a
linear diameter, $D = 640$ pc, and the HI mass is $6.4 \times 10^6$ \Msol \, 
(see Table 1). Other giant structures at the north and northwest of 
the ``main supershell'' show no evidences of regular shell expansion.

The ``main supershell'' was cut in 12 directions. We did not find any 
indication of regular expansion of the ``main supershell''
in any of the cuts. Only local gas perturbations were identified 
in some sectors of the ``main supershell'' (e.g. at the positions
corresponding to 115$^{\prime \prime}$ - 200$^{\prime \prime}$ of cut 2,  
135$^{\prime \prime}$ - 180$^{\prime \prime}$ of cut 3, 
430$^{\prime \prime}$ - 630$^{\prime \prime}$ of cut 5 and 
90$^{\prime \prime}$ - 210$^{\prime \prime}$ of cut 7).
The observed local velocities are in the range 215 - 245 km s$^{-1}$.
The best example of the local high velocity flow is observed in the 
vicinity of the WO star which is located at the southeast wall of the ``main 
supershell'' (see Lozinskaya et al., 2001). In this region we 
probably observe the blow-out of the WO stellar wind from 
the shell into the low density bubble interior. The local velocities 
observed in this region (see Figure 3b, at the positions corresponding to
190$^{\prime \prime}$ - 300$^{\prime \prime}$ along the cut 1) 
fall into the range -205 km~s$^{-1}$ - -250 km~s$^{-1}$.
We believe that other arc-shaped structures seen in the inner part 
of Segment 17 result from the collective feedback given 
by the associations 26, 34, 36, 37, 38, 42, 45 of Borissova et al. (2004),
which are located near the ridge of the ``main supershell''. The expansion 
velocities of the local arcs are less than or equal to 25 km s$^{-1}$.

The absence of regular expansion and the large thickness of the 
``main supershell'' led us 
to conclude that if this structure results from the combined action of
the embedded groups of massive stars, then the main supershell has 
already reached its standing position and is now confined by the
external gas pressure (see Koo \& McKee, 1992).

\section{The multiple SNe scenario}

In the standard SNe scenario, the shell expansion is supported by a high
thermal pressure provided by multiple supernova explosions and stellar 
winds from massive stars recently formed inside the superbubble volume.
In the case of an homogeneous ISM and a constant input rate of mechanical 
energy, the time evolution of the radius, $R$, the expansion 
velocity, $V$, and the mass, $M$, of the shell are given by 
(see Bisnovatyi-Kogan \& Silich, 1995, and references therein):
\begin{eqnarray}
      \label{eq.1a}
R & = & \left(\frac{125}{154 \pi}\right)^{1/5}
         \left(\frac{L_{mech}}{\rho_{ISM}}\right)^{1/5} t^{3/5}  \\
\label{eq.1b}
V  & = & \frac{3}{5} \frac{R}{t} \\
\label{eq.1c}
M  & = & \frac{4 \pi}{3} R^3 \rho_{ISM} \,; 
\end{eqnarray}
where $t$ is the bubble age, $L_{mech}$ is the rate of the mechanical energy
deposited by SNe explosions and stellar winds and $\rho_{ISM}$ is the
density of the ISM. Eliminating the evolutionary time $t$ and the ISM density 
$\rho_{ISM}$ from equations (\ref{eq.1a}) - (\ref{eq.1c}), 
one can find the mechanical power, $L_{mech}$, required to create a
shell with radius $R$, mass $M$ and expansion velocity $V$:
\begin{equation}
      \label{eq.2}
L_{mech} = \frac{77}{18} \frac{M V^3}{R} = 2.8 \times 10^{38} \,
           \frac{M_7 V_{10}^3}{R_{100}} \, \, erg \, s^{-1} \,,
\end{equation}
where $M_7$ is the mass of the shell in units of $10^7$\Msol,
$V_{10}$ is the expansion velocity in units of 10 km s$^{-1}$
and $R_{100}$ is the radius of the shell in units of 100 pc.

The transition to the pressure-confined stage occurs when the velocity
of the shock approaches the ambient gas turbulent speed, $V = \sigma_{ISM}$. 
At that time the shell begins to expand in both directions from
the stand-off radius, $R_{stand}$, with the local sound speed, 
$c_{sh}$. This may be consistent with the irregular boundary of 
the main supershell and with the standard scenario. If this is the
case, it takes the shell $\sim 0.5 \Delta R / c_{sh} \ge 10^7$ yr to 
reach the observed thickness. Thus the detected configuration may be 
observed for a good fraction ($\sim 1/4$) of the total evolutionary 
time, $t$. The stand-off radius is $R_{inner} < R_{stand} < R_{out}$, 
where $R_{inner}$ is the inner and $R_{out}$ is the outer
radius of the shell. We assume that 
\begin{equation}
      \label{eq.3}
R_{stand} = R_{inner} + \frac{\Delta R}{2} = 600 \, pc \,,
\end{equation}
where $\Delta R \approx 200$~pc is the mean thickness of the shell. 
For the rate of mechanical energy we then obtain:
$L_{mech} \approx 1.9 \times 10^{38}$ erg s$^{-1}$. 
This energy input rate requires a coeval star cluster
of approximately $10^4$\Msol \, with a Salpeter initial mass function 
with a lower and upper mass limits of 1\Msol \, and 100\Msol .
We do not see stellar remnants of such massive clusters in the field
of view of the ``main supershell''. On the contrary, OB-associations found 
inside the ring have a wide range of ages (see Table 2) that indicates
that in the region encompassed by the main supershell the star
formation lasts for more than 30 Myr. This suggests that a continuous 
star formation rate would be a better approximation for  
the star formation activity that occurred in this region.
 
In this case the rate of mechanical energy cannot be approximated by
a constant value during the whole evolution. In such a scenario the maximum 
value is reached $\sim 30$~Myr after the beginning of star 
formation (see Leitherer \& Heckman, 1995). 
On the other hand the thermal pressure of the interstellar gas also reduces
the shell radius below the standard value (Oey \& Garc\'{\i}a-Segura,
2004). The calculated energy may be also affected by the deviation of 
the ISM density from a homogeneous distribution.

To take into consideration these effects and to obtain a realistic
estimate for a SFR that is required to build up a shell of
similar to the observed structure, we have provided several numerical
runs using for calculations our 2.5D Lagrangian scheme based on 
the thin layer approximation (see Bisnovatyi-Kogan \& Silich 1995; 
Silich \& Tenorio-Tagle 1998). As an input model we adopt an 
exponential galactic disk with a constant velocity dispersion 
$\sigma = 10.2$~km~s$^{-1}$,
\begin{equation}
      \label{eq.4}
\rho_{ISM} = \rho_0 \exp{(-|z|/H_z)} \,,
\end{equation}
where $\rho_0$ is the midplane ISM density. It was assumed that the disk 
is embedded in a low density gaseous halo whose density was taken to be 
$n_{halo} = 10^{-5}$ cm$^{-3}$. We neglected the small shell distortion 
(Palou\v s et al., 1990; Silich, 1992) provided by the slow galactic rotation.
The Z-component of the gravitational field was calculated from the 
equilibrium conditions. The mass and energy deposition rates 
were calculated from our simplified starburst model (see Silich et al. 
2002) assuming that the SFR remains constant until the shell waist 
attains the observed radius. The SFR and the midplane ISM density 
$\rho_0$ were then calculated by iterations from the condition that the 
expansion velocity of the shell waist becomes equal to the ISM turbulent 
velocity and the calculated mass of the shell approaches the observed 
value when the radius of the shell waist reaches 600~pc.

The calculations have been provided for two different (thin and thick)
galactic disk models with the characteristic scale-heights 
of the exponential disk $H_z = 200$ pc and $H_z = 500$ pc, respectively.
Figure 4 presents the evolutionary sequence of superbubble shapes for
these two cases. 
\begin{figure*}
\vspace{20.0cm}
\centering
\includegraphics{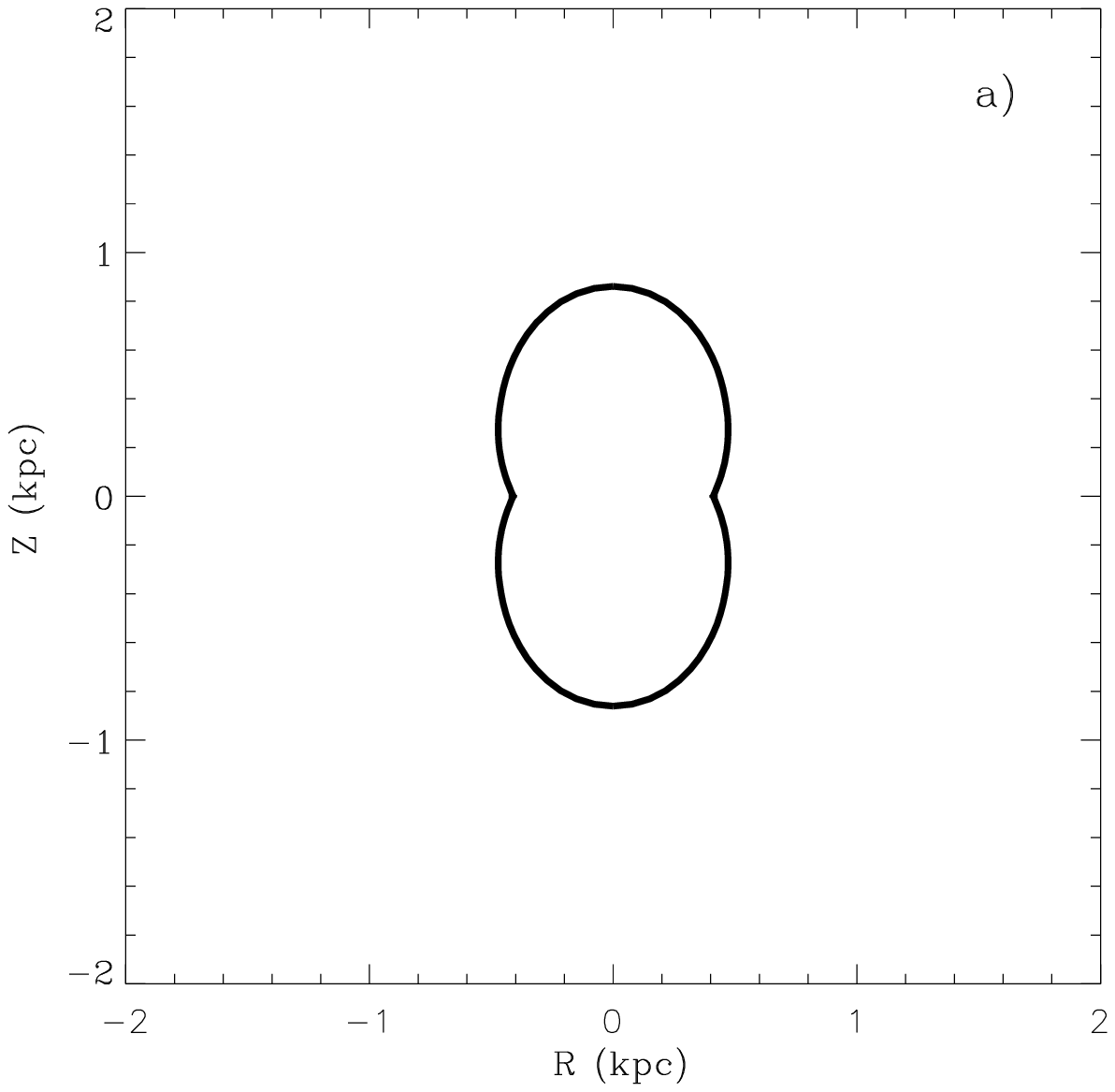}
\includegraphics{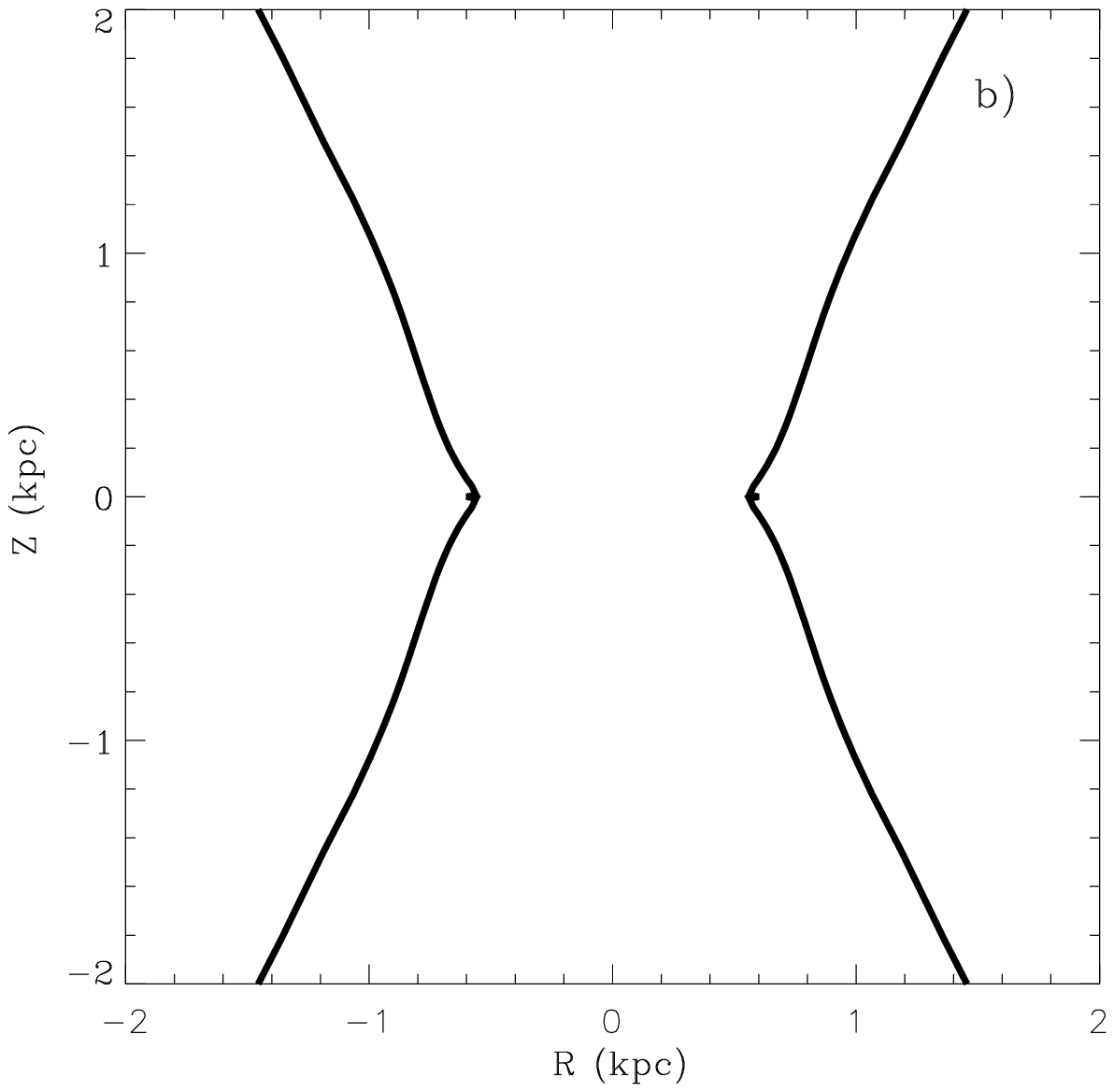}
\includegraphics{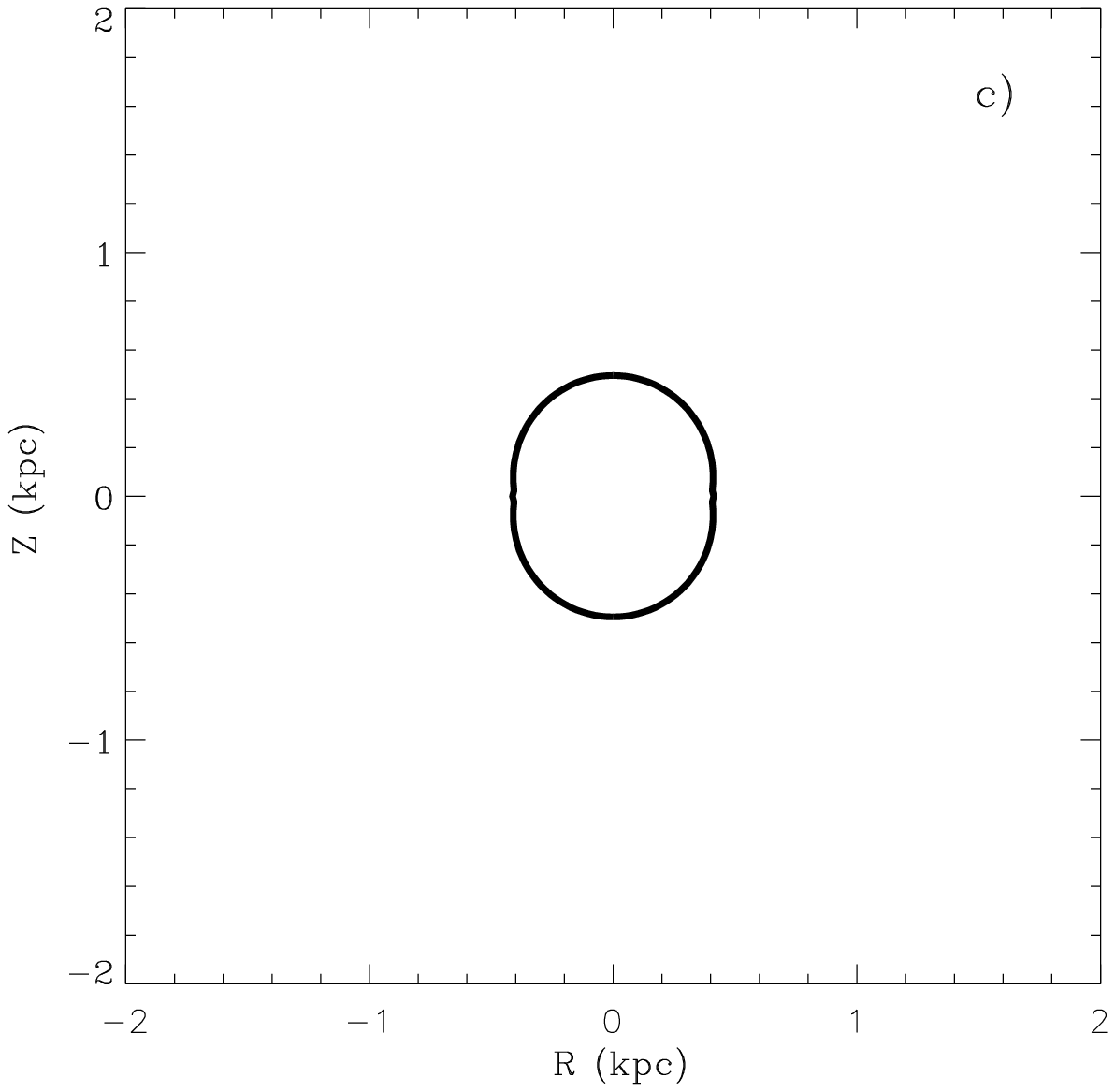}
\includegraphics{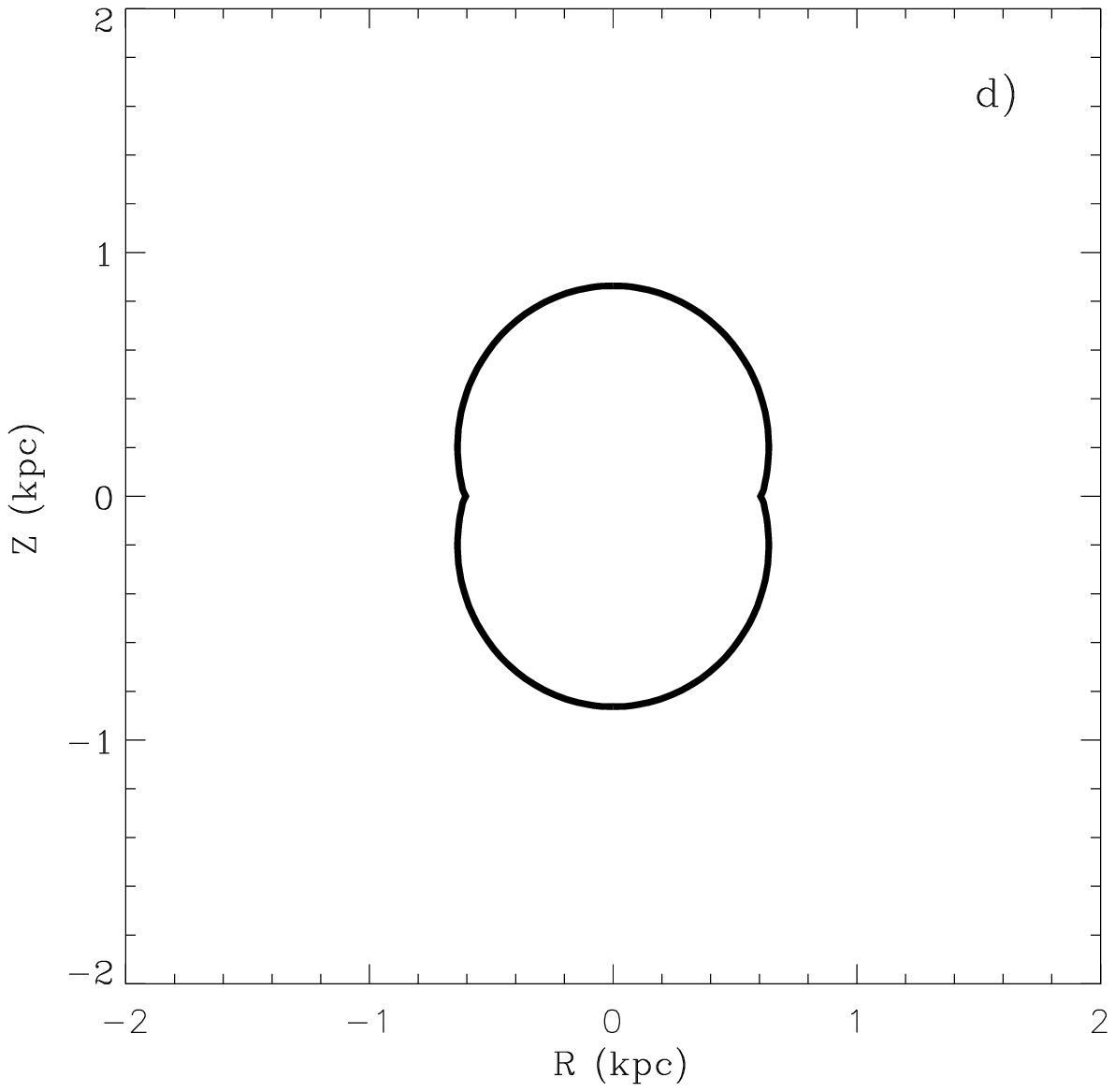}
\caption{The shape of the supershell. Panels a), b) and c), d) present 
the results of the calculations for the thin ($H_z = 200$ pc) and 
thick ($H_z = 500$ pc) disk models, respectively. The ages of the shells 
in panels a) and c) are similar ($\approx 20$Myr). Panels b) and d)
present supershells at the moment when they reach their standing positions 
at the midplane of the galaxy.}
\label{fig4}
\end{figure*}

In both cases the bubbles grow faster in the direction of the steepest 
density gradient. In the case of the thin disk, the superbubble already
reaches a typical 
hour-glass form after 20 Myr (Figure 4a). Soon the shell accelerates 
rapidly into the low-density galactic halo and the top of the shell is
disrupted under the action of Rayleigh-Taylor instabilities.
At this age ($\approx 20$~Myr), the top of the shell reaches approximately 
three characteristic Z-heights of the interstellar gas distribution.
The remnant is then able to ``blow out'', driving  its high temperature 
gas between the fragments into the halo of the host galaxy 
(Silich \& Tenorio-Tagle, 2001), and to establish the cone-like 
structure seen in Figure 4b. The bubble then loses its thermal
pressure and the waist of the shell reaches its standing position 
at the age $t \approx 34$ Myr. 
Despite the large size of the cone, the bulk of
the swept-up material is located within a thin cylindrical segment around
the base of the cone. In this particular case, about 50\% of the mass of
the remnant is concentrated inside a 400 pc (200 pc above and 200 pc 
below the midplane of the galaxy) segment.  

To fit the observed parameters of the main supershell the midplane gas 
number density should be $n_0 \approx 2.8$ cm$^{-3}$ and the star
formation rate SFR$ \approx 7.5 \times 10^{-3}$\Msol \, yr$^{-1}$.
This value is more then an order of magnitude larger than the local 
star formation rate, $SFR_{local} \approx (3-4) 
\times 10^{-4}$ \Msol yr$^{-1}$, derived from the analysis of
our $H{\alpha}$ data and stellar population of the embedded
OB-associations, and approximately three times larger than the average 
star formation rate obtained by Cole et al. (1999) for the entire
galaxy. Thus the thin-disk model is completely inconsistent with the the 
standard SNe scenario.

In the thick disk case the shell remains almost round after 20 Myr of 
expansion. Even at the end of the calculations, at  
$t \approx 37$Myr, the shell is only slightly distorted by the ISM 
density gradient (see Figure 4d). The Rayleigh-Taylor instabilities 
never disrupt the top of the shell and the hot gas remains bound
within the cylindrically-shaped remnant slowly expanding in the
Z-direction. To fit the observed parameters (the stand-off radius
and the mass of the shell), the mid-plane density of the disk should 
be $\approx 1.4$ cm$^{-3}$ and the star formation 
rate must be about $2.3 \times 10^{-3}$ \Msol \, yr$^{-1}$. 
This value significantly ($\sim 5$ times) exceeds our limit to the local 
star formation rate and is only slightly below the mean star formation rate
obtained for the entire galaxy. There are also no indications on a massive 
single stellar cluster that may support shell expansion down to the 
observed radius. This leads us to the conclusion that 
multiple SNe hypothesis is inconsistent with the observed parameters 
of the ``main supershell'' also in the case of the thick disk ISM 
distribution. 
Some different mechanism (e.g. the outward force initiated by the 
radiation pressure of the field stars, Elmegreen \& Chiang, 1982) 
is required to build up the observed structure.

\section{Conclusions}

Here we presented the results of our high resolution study of 
neutral hydrogen distribution and kinematics in a Local Group 
member, the dwarf irregular galaxy 
IC 1613 and compared them to our previous examination of the stellar
population and ionized component of the ISM in this galaxy. The
analysis of the 21 cm line data cube reveals a complicated, highly
inhomogeneous structure of the interstellar medium with a number of HI 
holes, shells and arc-like structures of different sizes and expansion 
velocities.

We found that smaller shells with diameters of a few hundred parsecs
tend to expand with velocities 10 km s$^{-1} \le V \le 20$
km s$^{-1}$ whereas the largest structures do not present evidences
of global expansion. The only exception is the large incomplete shell
with mass $\sim 6.4 \times 10^6$\Msol \, and expansion velocity,
$V \approx 13$ km s$^{-1}$ at the northwest part of the galaxy.
The best example of the large standing structure is an HI hole
1 kpc in diameter surrounded by a $2.8 \times 10^7$\Msol \, neutral 
hydrogen ring centered at RA(2000)=1h 4m 52s, D(2000)=2deg 07$^{\prime}$.

There are many OB-associations of different ages inside the hole whose
energetics potentially may be responsible for the development of this
structure. To examine this hypothesis we have
compared the energy required by the multiple supernovae model with that
inferred from the star formation rate derived from the
analysis of the embedded stellar population and H$_{\alpha}$ emission.
We found that parameters of the observed structure and the level of the
star formation activity detected inside the region encompassed by the
contour of the ``main supershell'' are inconsistent with the SNe 
hypothesis. This indicates that, as it has been revealed in the past 
for the largest 
HI holes in HoII and LMC (Rhode et al. 1999; Broun et al. 1997),
some different mechanism is required to build up the largest kpc-size
HI structure that is detected in IC 1613.
 
\begin{acknowledgements}
The authors thank William Wall and Guillermo Tenorio-Tagle for careful
reading of the manuscript and suggestions and our referee,
Dr. A. Burkert, for his critical comments. We also greatly appreciate 
help from the stuff of The National Radio Astronomy Observatory 
(NRAO) where the observations were performed. NRAO is part of the 
National Science Foundation (USA) and is operated by the Association 
of Universities Inc. under a contract with the NSF. This study was 
supported by CONACYT M\'exico  research grants 47534-F and 40095-F, 
by DGAPA-UNAM grant IN120802, by Russian Foundation for Basic 
Research (projects 02-02-06048 and 04-02-16042). A.V.M. also wishes 
to thank the Russian Science Support Foundation.
\end{acknowledgements}

\newpage

\end{document}